\documentclass[fleqn,10pt]{wlscirep}
\usepackage[utf8]{inputenc}
\usepackage[T1]{fontenc}

\usepackage[version=4]{mhchem}
\usepackage{siunitx}
\usepackage{bm}
\usepackage{xcolor}

\DeclareSIUnit\Molar{M}

\newcommand{\bs}{{\boldsymbol{\sigma}}}
\newcommand{\bt}{{\boldsymbol{\theta}}}

\title{Unsupervised Bayesian Ising Approximation for revealing the neural dictionary in songbirds}

\author[1,2]{Dami\'{a}n G. Hern\'{a}ndez}
\author[3]{Samuel J. Sober}
\author[2,3,4]{Ilya Nemenman}
\affil[1]{Department of Medical Physics, Centro At\'{o}mico Bariloche and Instituto Balseiro, Bariloche 8400, Argentina}
\affil[2]{Department of Physics, Emory University, Atlanta, GA 30322, USA}
\affil[3]{Department of Biology, Emory University, Atlanta, GA 30322, USA}
\affil[4]{Initiative in Theory and Modeling of Living Systems, Atlanta, GA 30322, USA}

\begin{abstract}
The problem of deciphering how low-level patterns (action potentials in the brain, amino acids in a protein, etc.) drive high-level biological features (sensorimotor behavior, enzymatic function) represents the central challenge of quantitative biology. The lack of general methods for doing so from the size of datasets that can be collected experimentally severely limits our understanding of the biological world. For example, in neuroscience, some sensory and motor codes have been shown to consist of precisely timed multi-spike patterns. However, the combinatorial complexity of such pattern codes have precluded development of methods for their comprehensive analysis. Thus, just as it is hard to predict a protein’s function based on its  sequence, we still do not understand how to accurately predict an organism's behavior based on neural activity. Here we derive a method for solving this class of problems. We demonstrate its utility in an application to neural data, detecting precisely timed spike patterns that code for specific motor behaviors in a songbird vocal system. Our  method detects such codewords with an arbitrary number of spikes, does so from small data sets, and accounts for dependencies in occurrences of codewords. Detecting such dictionaries of important spike patterns – rather than merely identifying the timescale on which such patterns exist, as in some prior approaches – opens the door for understanding fine motor control and the neural bases of sensorimotor learning in animals. For example, for the first time, we identify differences in encoding  motor exploration versus typical behavior.  Crucially, our method can be used not only for analysis of neural systems, but also for understanding the structure of correlations in other  biological and nonbiological datasets.
\end{abstract}
\begin{document}

\flushbottom
\maketitle

\thispagestyle{empty}

\section*{Introduction}

\begin{figure}
   \includegraphics[width=0.9\textwidth]{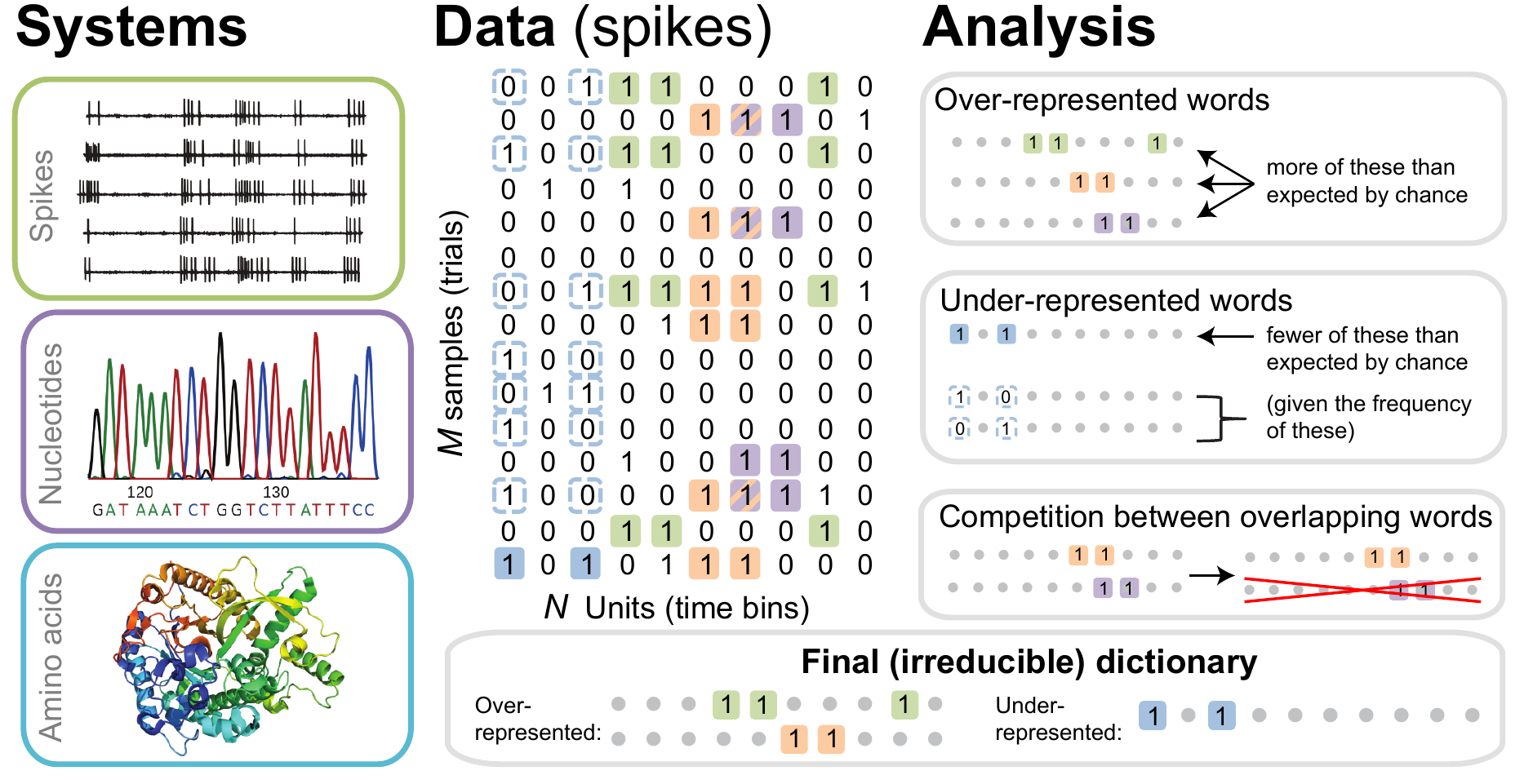}
\caption{\footnotesize{{\bf The dictionary reconstruction problem.} In many biological systems, such as understanding the neural code, identifying protein-DNA binding sites, or predicting 3-D protein structures, we need to infer dictionaries --- the sets of statistically over- or under-represented features in the datasets, which we refer to as words in the dictionary. To do so, we represent the data as a matrix of binary activities of $N$ biological units (spike/no spike, presence/absence of a mutation, etc.), and view $M$ different experimental instantiations as samples from an underlying stationary probability distribution. We then use the uBIA method to identify the significant words in the data. Specifically, uBIA systematically searches for combinatorial activity patterns that are over- or under-represented compared to their expectation given the marginal activity of the individual units. If multiple similar patterns can (partially) explain the same statistical regularities in the data, they compete with each other for importance, resulting in an irreducible dictionary of significant codewords. In different biological problems, such dictionaries can represent neural control words, DNA binding motifs, or conserved patterns of amino acids that must be neighbors in the 3-D protein structure.}\label{f0}}
\end{figure}

One of the goals of modern high-throughput biology is to generate predictive models of interaction networks, from interactions among individual biological molecules \cite{marks2011protein} to the encoding of information by networks of neurons in the brain \cite{schneidman2006weak}. To be able to make predictions about activity across networks requires one to accurately approximate---or build a {\em model} of---their joint probability distribution, such as the distribution of joint firing patterns in neural populations or the distribution of co-occurring mutations in proteins of the same family. To successfully generalize and to improve interpretability, models should contain as few as possible terms. Thus constructing a model requires detecting {\em relevant} features in the data: namely, the smallest possible set of spike patterns or nucleotide sequences that capture the most correlations among the network components. By analogy with the human language, where words are strongly correlated, co-occurring combinations of letters, we refer to the problem of detecting features that succinctly describe correlations in a data set as the problem of {\em dictionary reconstruction}, see Fig.~\ref{f0}.

In recent years, the problem of dictionary reconstruction has been addressed under different names for a variety of biological contexts \cite{natale2017reverse} including gene expression networks \cite{margolin2006aracne,lezon2006using}, protein structure, protein-protein interactions \cite{marks2011protein,weigt2011, bitbol2016inferring,halabi2009protein}, the structure of regulatory DNA  \cite{otwinowski2013genotype}, distribution of antibodies and pathogenic sequences  \cite{mora2010maximum,ferguson2013computational}, species abundance \cite{tikhonov2015interpreting}, and collective behaviors \cite{bialek2012statistical, couzin2003self, lukeman2010inferring, kelley2013emergent, perez2011collective}. The efforts to identify interactions in neural activity have been particularly plentiful \cite{Stevens:1995ir, schneidman2006weak, pillow2008spatio, bassett2017network, williams2019discovering}. The diversity of biological applications notwithstanding, most of these attempts have relied on similar mathematical constructs, and most have suffered from the same limitations. First, unlike in classical statistics and traditional quantitative model building, where the number of observations, $M$, usually vastly exceeds the number of unknowns to be estimated, $K$, $K/M\ll1$, modern biological data often has $M\gg1$, but also $K/M\gg1$. Indeed, because of, for example, protein allostery, recurrent connections within neural populations, or coupling to global stimuli, biological systems are rarely limited to local interactions only \cite{Schwab:2013, merchan2016sufficiency, nemenman:2017}, so that the number of pairwise interactions among $N$ variables is $K \sim N^2$, and the number of all higher order interactions among them is $K\sim 2^N$. Put differently, words in biological dictionaries can be of an arbitrary length, and spelling rules may involve many letters simultaneously, some of which are far away from each other. Because of this, reconstruction of biological dictionaries from data sets of realistic sizes requires assumptions and simplifications about the structure of possible biological correlations, and will not be possible by brute force. The second problem is that, as in human languages, biological dictionaries have redundancies: there are synonyms and words that share roots. For example, a set of gene expressions may be correlated not because the genes interact directly, but because they are related to some other genes that do. Similarly, a certain pattern of neural activity may be statistically over- or under-represented not on its own, but because it is a subset or a superset of another, more important, pattern. Identifying {\em irreducible words}---the root forms of biological dictionaries---is therefore harder than detecting all correlations while also being crucial to fully understanding biological systems. Altogether, these complications make it impossible to use off-the-shelf methods for reconstructing combinatorially complex dictionaries from datasets of realistic sizes. 

In this work, we propose a new method for reconstructing complex biological dictionaries from relatively small datasets, as few as $M \sim 10^2\dots10^3$ samples of the joint activity. We impose no limitation on the structure of the words that can enter the dictionary --- neither their length nor their rules of spelling --- beyond the obvious limitation that (i)  words that do not happen in the data cannot be detected, and (ii) that data contain few samples of many words, rather than of just a few that repeat many times. Additionally, we address the problem of irreducibility, making the inferred dictionaries compact, non-redundant, and easier to comprehend. The main realization that allows this progress is that instead of approximating the entire joint probability distribution of a system's states and hence answering {\em how} specific significant words matter, we can focus on a more restricted, and hence simpler, question: {\em which} specific words contribute to the dictionary. We answer this question in the language of Bayesian inference and statistical mechanics by developing an unsupervised version of the Bayesian Ising Approximation \cite{fisher2015bayesian} and by merging it with the {\em reliable interactions} model \cite{Ganmor:2011epa}.

To illustrate the capabilities of our approach, we develop it in the context of a specific biological system: recordings from single neurons in brain area RA (the robust nucleus of the arcopallium) in the Bengalese finch, a songbird. Neurons communicate with each other using patterns of  action potentials (spikes), which encode sensory information and motor commands, and hence behavior. Reconstructing the neural dictionary, and specifically detecting irreducible patterns of neural activity that correlate with (or ``encode'') sensory stimuli or motor behaviors --- which we hereafter call {\em codewords} --- has been a key problem in computational neuroscience for decades \cite{Stevens:1995ir}. It is now known that both in the sensory \cite{berry1997structure, strong1998entropy, reinagel2000temporal, arabzadeh2006deciphering, rokem2006spike, nemenman2008neural, lawhern2011spike, fairhall2012information} and in the motor systems \cite{tang2014millisecond, Srivastava31012017,sober2018millisecond, Putney:2019} the timing of neural action potentials (spikes) in multispike patterns, down to millisecond resolution, can contribute to the encoding of  sensory or motor information. Such dictionaries that involve long sequences of neural activities (or incorporate multiple neurons) at high temporal resolution are both complex and severely undersampled. Specifically, even though the songbird datasets considered here are large by neurophysiological standards, they are too small for most  existing analysis approaches for modeling neural activity to detect the codewords. This motivates the general inference problem we address here.

Understanding the neural-motor dictionary answers important questions about vertebrate motor control which could not be addressed by previous techniques (see {\em Online Methods}). Specifically, while it is known that various features of the complex vocal behaviors are encoded by millisecond-scale firing patterns \cite{tang2014millisecond}, here for the first time we can identify which specific patterns most strongly predict behavioral variations. Further, we show that dictionaries of individual neurons are rather large and quite variable, so that neurons speak different languages, which nonetheless share some universal features. Intriguingly, we detect that codewords that predict large, exploratory deviations in vocal acoustics are statistically different from those that predict typical behaviors. Collectively, these findings pave the way for development of future theories of the structure of these dictionaries, of how they are formed during development, how they adapt to different contexts, and how motor biophysics translates them into specific movements. More generally, these findings open gates for using our method in other biological domains, where reconstruction of feature dictionaries is equally important.
\begin{figure}
   \includegraphics[width=0.9\textwidth]{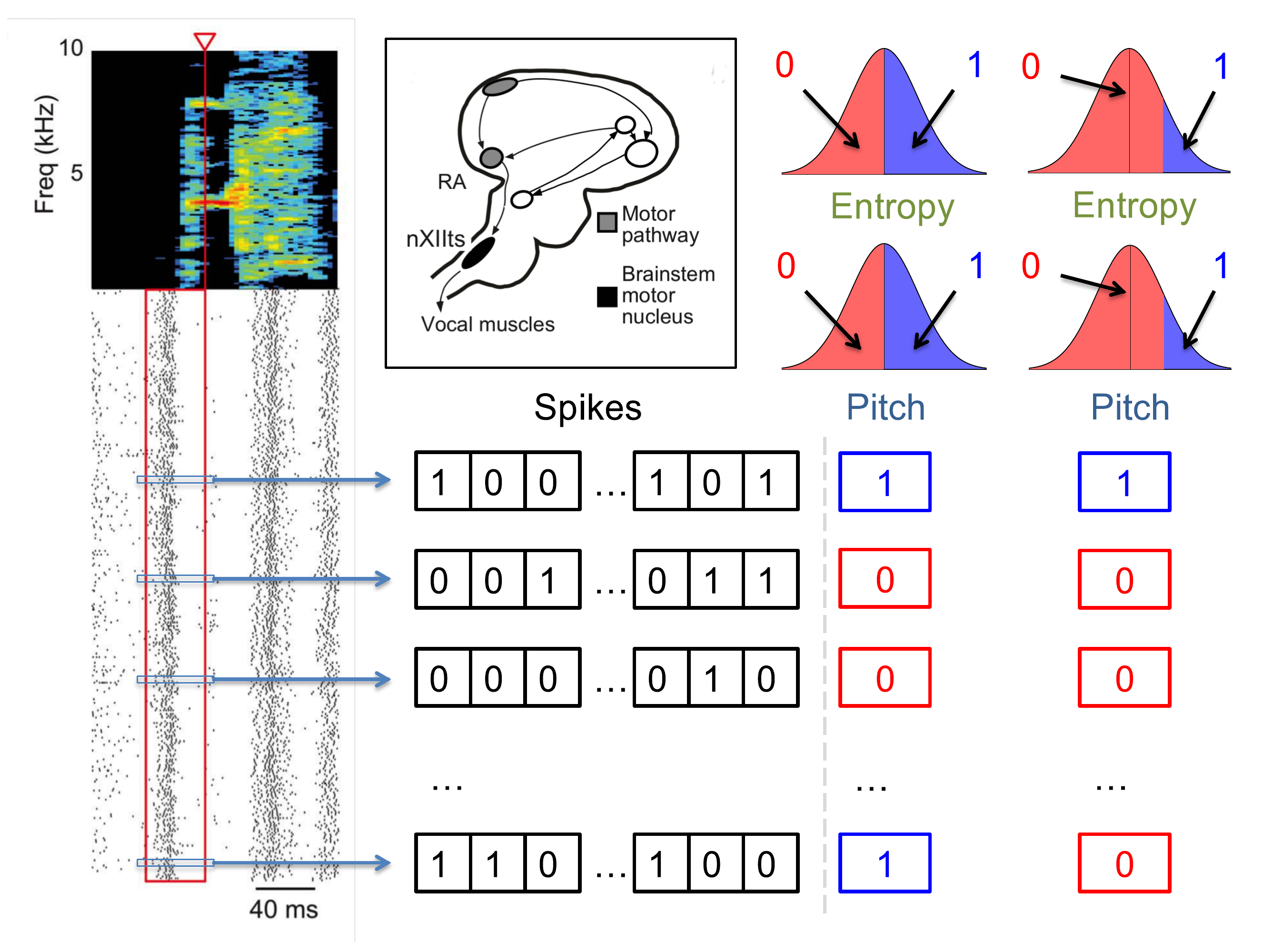}
\caption{\footnotesize{{\bf Quantification of the neural activity and the behavior.} A spectrogram of a single song syllable in top-left corner shows the acoustic power (color scale) at different frequencies as a function of time. Each tick mark (bottom-left) represents one spike and each row represents one instantiation of the syllable. We analyze spikes produced in a 40 ms premotor window (red box) prior to the time when acoustic features were measured (red arrowhead). These spikes were binarized as $0$ (no spike) or $1$ (at least one spike) in 2 ms bins, totaling $20$ time bins. The different acoustic features (pitch, amplitude, spectral entropy) were also binarized. For different analyses in this paper, 0/1 can denote the behaviors feature that is below/above or above/below its median, or as not belonging/belonging to a specific percentile interval. The inset shows the area RA within the song pathway, two synapses away from the vocal muscles, from which these data were recorded.}} \label{f1}
\end{figure}

\section*{Results}

\subsection*{The neuro-motor code reconstruction problem}

Owing to their complex and tightly regulated vocal behavior and experimentally accessible nervous system, songbirds provide an ideal model system for investigating the neural dictionaries underlying complex motor behaviors \cite{Doupe, kuebrich2015variations}. We recorded from individual neurons in the motor cortical area RA of Bengalese finches during spontaneous singing, while at the same time quantifying acoustic feature of song, specifically the fundamental frequency ("pitch"), amplitude, and spectral entropy of individual vocal gestures, or "syllables'', as described previously \cite{sober2008central,tang2014millisecond,wohlgemuth2010}. The data sets are sufficiently large to allow reconstruction of the dictionaries: we have $49$ data sets --- spanning $4$ birds, $30$ neurons and sometimes multiple song syllables for each neuron --- for which we observed at least $200$ instances of the behavior and the associated neural activity, which we estimate below to be the lower threshold for a sufficient statistical power.

To represent analysis of this motor code as a dictionary reconstruction problem, we binarize the recorded spiking time series so that $\sigma_t=(0,1)$ indicates the absence or presence of a spike in a time slice of $[(t-1)\Delta t, t\Delta t)$, see Fig.~\ref{f1}. Thus each time interval is represented by a binary variable, and interactions among these patterns are described by over-represented or under-represented sequences of zeros and ones in the data. Using a complementary information-theoretic analysis, Tang et al.\ \cite{tang2014millisecond} showed that the mutual information between the neural spike train and various features of song acoustics peaks at $\Delta t =1\dots2$ ms. Thus studying precise timing codes means that we focus on $\Delta t=2$ ms (our datasets are not large enough to explore smaller $\Delta t$) as discussed previously in  \cite{tang2014millisecond}.  Detection of statistically significant codewords at this temporal resolution would both re-confirm  that this neural motor code is timing based, consistent with previous analyses \cite{tang2014millisecond}, as well as for the first time revealing the specific patterns that most strongly predict behavior. We focus on neural time series of length $T=40$ ms duration preceding a certain acoustic syllable, which includes the approximate premotor delay with which neurons and muscles influence behavior \cite{tang2014millisecond}. Thus the index $t$ runs between 1 and $N=T/\Delta t=20$.

Since we are interested in codewords that relate neural activity and  behavior, we similarly binarize the motor output \cite{tang2014millisecond}, denoting by $0$ or $1$ different binary characteristics of the behavior, such as pitch being either below or above its median value, or outside or inside a certain range, see Fig.~\ref{f1}. We treat the behavioral variable as the 0'th component of the neuro-behavioral activity network, which then has  $N=21$ binary variables,  $\bm{\sigma}=\{\sigma_i\}_0^{N}$. Finding the codewords and reconstructing the neural-behavioral dictionary is then equivalent to detecting significantly over- or under-represented patterns in the probability distribution $P(\bm{\sigma})$, and specifically those patterns that are overrepresented together with the behavioral bit. Note that $2^N=2^{21}\approx 2\cdot 10^6$, which is much greater than $M\sim 100\dots1000$ observations of the activity that we can record, illustrating the complexity of the problem. In fact, similar to the famous birthday problem, one expects a substantial number of repeating samples of the activity --- and hence the ability to detect statistically over- and under-represented binary words -- only when $M\sim \sqrt{2^N}$, which is what limits the statistical power of any dictionary reconstruction method.    

\subsection*{The unsupervised BIA method (uBIA)}
To reconstruct the neural-motor dictionaries and detect codewords that predict specific behaviors, we develop the unsupervised  Bayesian Ising Approximation (uBIA) method based on the Bayesian Ising Approximation for detection of significant features in regression problems \cite{fisher2015bayesian}. For this, we write the probability distribution $P(\bm{\sigma})$ as
\begin{multline}
  \log P(\bm{\sigma}|\vec\theta)=-\log Z +\sum_{i=0}^{N} \theta_i\sigma_i +\sum_{j\ge i}^{N}
               \theta_{ij}\sigma_i\sigma_j+\sum_{k\ge j\ge i}^{N}
               \theta_{ijk}\sigma_i\sigma_j\sigma_k+\dots+\theta_{0\dots N}\sigma_0\times\dots\times
               \sigma_{N}\label{loglinear}= -\log Z +\sum_{\mu}\theta_\mu\prod_{i\in V_\mu}\sigma_i,
\end{multline}
where $Z$ is the normalization coefficient \cite{amari2001information}. We use the notation such that $V_\mu$ is a nonempty subset of indexes $i\in[0,N]$, and $\mu=[1,2^{N+1}-1]$ is the subset number. Then $\{\theta_\mu\}=\vec\theta$ are coefficients in the log-linear model in front of the corresponding product of binary $\sigma$s. In other words, $V_\mu$ denotes a specific combination of the behavior and / or times when the neuron is active. If $\theta_\mu$ is nonzero for a term $\prod_{i\in V_\mu} \sigma_i$, where $i=0$ (the response variable) is in $V_\mu$, then this specific spike word is correlated with the motor output, and is a significant codeword in the neural code, see Fig.~\ref{f0}. Finding nonzero $\theta_\mu$s is then equivalent to identifying {\em which} codewords matter and should be included in the dictionary in Fig.~\ref{f0}, and inferring the exact values of $\theta_\mu$ tells {\em how} they matter.

A common alternative model of probability distributions uses  $x=2\sigma -1=\pm 1$ instead of $\sigma=(0,1)$. A third order term coupling, for example, $\sigma_i\sigma_j\sigma_k$ represents a combination of first, second, and third order terms in the corresponding $x$s, and vice versa. Thus which words are codewords may depend on the parameterization used, but the longest codewords and nonoverlapping groups of codewords remain the same in both parameterizations. Our choice of $\sigma$ vs $x$ is for a practical reason: a codeword in the $\sigma$ basis does not contribute to $P$ unless {\em all} of its constituent bins are nonzero. Thus since spikes are rare, we do not need to consider contributions of very long words to the code.

We would like to investigate the neural dictionary systematically and to avoid pitfalls of other methods that arbitrarily truncate Eq.~({\ref{loglinear}}) at some order of interactions. For this, we define indicator variables $s_\mu=(0,1)$, $\mu=1,\dots,2^{N+1}-1$, which denote if a particular sequence of $\sigma_i=1$, $i\in V_\mu$, and $\sigma_i=0$, $i\not\in V_\mu$, ``matters'' (is a putative word in the dictionary), i.~e., it is either statistically significantly over- or under-represented in the data set compared to a null model (which we define later). In other words, we rewrite
\begin{equation}
  \log P(\bm{\sigma}|\vec\theta)= -\log Z +\sum_{\mu}\theta_\mu s_{\mu}\prod_{i\in V_\mu}\sigma_i.
\label{indicators}
\end{equation}
We then choose a prior on $\theta_\mu$ and on $s_\mu$. As mentioned above, detecting words that are severely anomalously represented is easy, and it is not our focus. Instead, since many neurons control the muscles and hence the behavioral output, we assume that individual neuron can only have a very weak effect on the motor behavior. In other words, typically $|\theta_\mu|\ll 1$. We thus work in the strong regularization limit and impose priors
\begin{equation} {\cal P}(\theta_\mu|s_\mu=1)\propto \exp
  \left[-\frac{1}{2}\epsilon{(\theta_\mu-\theta^*_\mu)^2}\right],\;\epsilon\ll1. \label{strongprior}
\end{equation}
Note that the prior distribution ${\cal P}(\theta_\mu|s_\mu=0)$ is irrelevant since, for $s_\mu=0$, $\theta_\mu$ does not contribute to $P(\bm{\sigma}|\vec\theta,\vec s)$.  We then choose $\bar\theta_\mu$ in a way such that the {\em a priori} and the empirical averages of individual $\sigma_i$s are equal, $\langle \sigma_i\rangle=\overline{\sigma_i}$ (we always use $\overline{\dots}$ and $\langle\dots\rangle$ to denote {\em a priori} and {\em a posteriori} expectations). This is equivalent to saying that we constrain our model to reproduce the firing rate of neurons and the frequency of the behavior. In typical problems, such marginal expectations are, indeed, well-known, and it is the higher order interaction terms, the  complex words in the dictionary, that make the reconstruction hard. Finally, we choose $P(s_\mu=1)=P(s_\mu=1)=0.5$, so that {\em a priori} a particular codeword has a fifty-fifty chance of being included in the neural dictionary.

Since we are only interested in whether a spike word is important in predicting the behavior and not how important it is, we integrate out all $\theta_\mu$ having observed $M$ samples of the $N+1$ dimensional vector $\bm{\sigma}$. We perform this integration to the quadratic order in $\epsilon$ to get the posterior probability of indicator variables (see {\em Online Methods}):
\begin{equation}
P_{\epsilon}(\mathbf{s}|\{\bm{\sigma}\}) =\frac{1}{\mathcal{Z}(\epsilon)} \exp  \left[ \epsilon \sum_{\mu} h_{\mu}(\{\bm{\sigma}\}) s_{\mu}+\epsilon^2 \sum_{\mu\nu} J_{\mu\nu}(\{\bm{\sigma}\}) s_{\mu}s_{\nu} \right].
\label{s2e1}
\end{equation}
This is of a familiar pairwise Ising form \cite{thompson2015mathematical}, with data-dependent magnetic fields $h_\mu$ and the couplings $J_{\mu\nu}$. Note that this Ising model has $2^{N+1}$ spins, replacing the model with $N+1$ spins with higher order interactions in Eq.~(\ref{loglinear}). This is naively a much harder problem. However, since most of the $2^{N+1}$ words do not appear in the actual data, and because of the $\epsilon^2$ in front of the pairwise coupling term, evaluating posterior expectations $\langle s_\mu\rangle$ for all word that actually occur is relatively easy (see {\em Online Methods}). Knowing  $\langle s_\mu\rangle$ allows us to tell which specific words should enter the neural-behavioral dictionary with a high {\em a posteriori} probability. To interpret Eq.~(\ref{s2e1}), we notice (see details in {\em Online Methods}) that the linear terms $h_{\mu}(\{\bm{\sigma}\})$ bias the indicator variables: those words that over- or under-occur (are anomalously represented) in the data relative to their expected frequency in the null model that matches $\langle \sigma_i\rangle$ will have $h_{\mu}>0$. In their turn, the couplings $J_{\mu\nu}$ are typically negative for those words that frequently co-occur, which includes words that at least partially overlap. Thus in this Ising model, words compete to explain certain correlations in data. Once a word has a large evidence (the $h_\mu$ term), it suppresses all other correlated words that explain the data in a weaker way, bringing the dictionary closer to the irreducible form. To verify that our analysis can, indeed, recover neural dictionaries and to set various adjustable parameters involved in the method, we successfully tested the approach on synthetic data that are statistically similar to our neural recordings (see {\em Online Methods}). In particular, for the data we have, we expect false discovery rate of $<1$ codeword per neuron, but the fraction of the words that we discover depends on the (unknown) typical values of $\theta$s.

\begin{figure}
\centering
\includegraphics[width=0.9\textwidth]{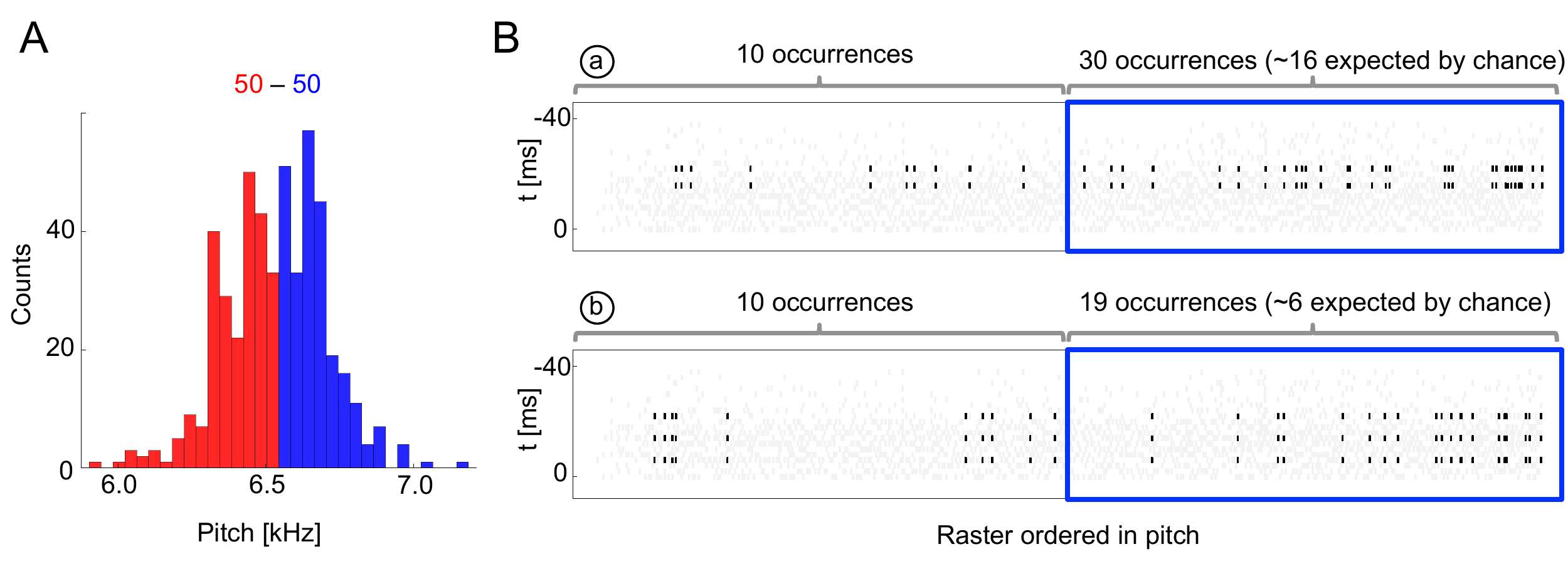}
\caption{\footnotesize{{\bf Sample multispike codewords.} {\bf A}: Probability distribution an acoustic parameter (fundamental frequency, or pitch). For this analysis, we consider the output to be $\sigma_0=1$ when the pitch is above median (blue), and zero otherwise (red). {\bf B}: Distribution of two sample codewords (a two-spike word in the top raster, and a three-spike word in the bottom raster) conditional on pitch. In each raster plot, a column represents 40 ms of the spiking activity preceding the syllable, with a grey tick denoting a spike. Every time a particular pattern is observed, its ticks are plotted in black. Note that these two spike words are codewords since they are overrepresented for above-median pitch (blue box) compared both to the null model based on the marginal expectation of individual spikes, and to the presence of the patterns in the low pitch region. Labels (a) and (b) identify these patterns in Fig.~\ref{f2}B.}}
\label{scatter}
\end{figure}

\subsection*{Statistical properties of neural motor codes}

\begin{figure}
\centering
\includegraphics[width=0.9\textwidth]{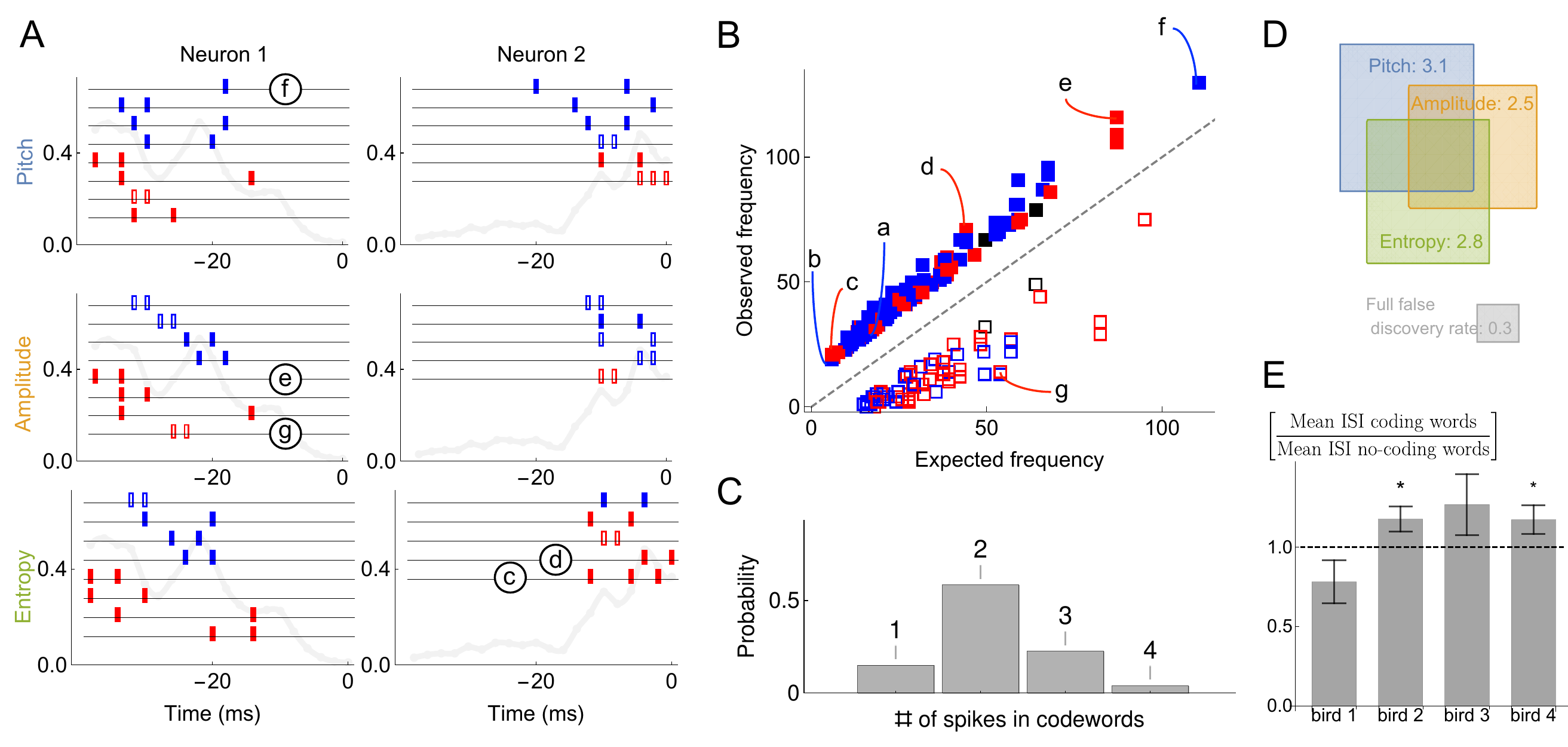}
\caption{\footnotesize{{\bf Statistical properties of neural dictionaries.} {\bf A}: Sample neural-behavioral dictionaries for two neurons from two different birds (columns) and for three different acoustic features of the song (rows: pitch, amplitude, and the spectral entropy). The light gray curve in the background and the vertical axis corresponds to the probability of neural firing in each 2 ms bin (the firing rate). The rectangular tics represents the timing of spikes in neural words that predict the acoustic features. For example, a two spike word with tics at points $t=i,j$ corresponds to the probability that the word $\mu=(i,j)$ is a codeword for the acoustic feature with a probability statistically significantly higher than $1/2$. Codewords for high (low) output, i.~e., $\sigma_0=1$ above (below) the median, are shown in blue (red). Full (empty) symbols correspond to over(under)-occurrence of the codeword-behavior combinations compared to the null model. Finally full (empty) black symbols represent words that over(under)-occur in the blue code and under(over)-occur in the red code. Words labeled (c)-(g) are also shown in (B). {\bf B}: Frequency of occurrence  of statistically significant codewords for different acoustic features in different neurons. Only first 200 codewords shown for clarity. Plotting conventions same as in (A), and letters label the same codewords as in (A) and in Fig.~\ref{scatter}B. 
{\bf C}: Proportion of $m$-spike codewords found in the dictionaries analyzed. An $m$-spike word corresponds to an $(m+1)$-dimensional word in the neural-behavioral dictionary. Most of the significant codewords have two or more spikes in them. {\bf D}: Mean number of significant codewords, averaged across all neurons and acoustic features. An average neuron has $5.6$ codewords in our dataset, of which $3.1$ code for the pitch, $2.5$ for the amplitude, and $2.8$ for the spectral entropy, with the number of words coding for pairs of features or for all three of them indicated by the overlap of rectangles in the Venn diagram. For comparison, our estimated false discovery rate is 0.3 words, so that only $\sim 0.3$ spurious words are expected to be discovered in each individual dictionary. We note that about a third of all analyzed dictionaries are empty, so that those that have words in them typically have more than illustrated here. {\bf E}: Mean inter-spike interval (ISI) for the codewords (spike words that code for behavior) vs.\ all spike words that are significantly over- or under-represented, but do not code for behavior. Averages in each of the four analyzed birds are shown, illustrating that the ISI statistics of the coding and non-coding words are different, but the differences themselves vary across the birds. Star denotes 95\% confidence. Other properties of the dictionaries (mean number of spikes in codewords, fraction of codewords shared by three vocal features, proportion of under/over-occurring codewords), do not differ statistically significantly across the birds.}}
\label{f2}
\end{figure}

Figure~\ref{scatter} illustrates the occurence of two specific codewords found by uBIA that encode high-pitch renditions of syllables. Note that these codewords are, indeed, overrepresented together with the high pitch vocalizations. Analyzing if a particular word is correlated with an acoustic feature is, of course, not hard. However, detecting words that should be tested, without a multiple hypothesis testing significance penalty is nontrivial. Thus the power of uBIA comes from being able to systematically analyze abundances of {\em combinatorially many} such spike  patterns, and further to identify which of them are {\em irreducibly} over- or under-represented. Figure~\ref{f2} illustrates statistical properties of entire neural-behavior dictionaries discovered by uBIA for different songbird premotor neurons and for three features of the acoustic behavior. While we reconstruct the dictionaries that include all irreducible words, including those that have only anomalous firing patterns, here we primarily focus on codewords, that is the words that relate behavior to the neural activity. We do the analysis twice, first for behavior binarized as $\sigma_0=1$ for the above-median acoustic features, and then for the below-median acoustic features. We then  combine the results. Note that the same pattern of spikes should not be simultaneously over- or under-represented when studying both the above and the below median codes, since the pattern cannot code for two mutually exclusive features. There were $0.7$ such codewords on average per dictionary. This is consistent with the expected false discovery rate of about 0.3 codewords per neuron for data sets of our size and statistical properties.

The most salient observation is that the inferred codewords consist of present or absent spikes in specific 2 ms time bins. This is consistent with previous analysis  \cite{tang2014millisecond}, which identified the same timescale for this dataset by analyzing the dependence of the mutual information between the activity and the behavior on the temporal resolution, but was unable to detect the specific words that carry the information. The second crucial observation is that most of codewords are composed of multiple spikes, representing an orthographically complex {\em pattern} timing code \cite{sober2018millisecond}, in contrast to single spike timing codes, such as in \cite{bialek1991reading}. Large number of codewords of 2 or more spikes (and thus 3 or more features, including the behavior itself) suggests that analyzing these dictionaries with the widely-used lower order MaxEnt or GLM methods that typically focus on lower-order statistics (see {\em Online Methods}) would miss their significant components. Our third crucial observation is that very few sub-words / super-words pairs occur in the dictionaries, cf.~Fig.~\ref{f2} (e.~g., the second codeword coding for entropy in neuron 2 in the panel A is a subword of the others). This indicates that uBIA fulfills its goal of rejecting multiple correlated explanations for the same data.

We quantify these observations as follows. In the $49$ different datasets, the average size of a dictionary is $14$. Of these words, on average $5.6$ include the behavioral feature and hence are {\em codewords}, cf.~Fig.~\ref{f2}(D). That there are so many specific temporally precise codewords suggests that the behaviorally-relevant spike timing patterns are the rule, rather than the exception, in this dataset. We found that $66\%$ of codewords are unique to one of the three analyzed acoustic features. This further quantifies the observation that some neurons in RA are {\em selective} for specific acoustic features, as noted previously in \cite{sober2008central}. Across all neurons and all acoustic features, only $15\%$ of codewords consist of a single spike (or absence of spike), while $58\%$, $23\%$, and $4\%$ consist of two, three, and four spikes respectively, cf.~Fig.~\ref{f2}(C) (we are likely missing many long codewords, especially with small $\theta$'s due to undersampling, see {\em Online methods}). This observation is consistent across all neurons and acoustic features, again indicating that coding by temporally precise spike patterns is a rule and not an exception.

At the same time, the observed dictionaries are quite variable across neurons and the production of particular song syllables. Codewords are built by stitching together multiple spikes or spike absences, and individual spikes occur at certain time points in the $(-40,0)$ ms window with different probabilities in different neurons and syllables (i.~e., the firing rate is both time and neuron dependent, cf.~Fig.~\ref{f2}(A), grey lines). Codewords are likely to occur where the probability of seeing a spike in a bin is $\sim50\%$, since these are the times that have more capacity to transmit information. Thus variability in firing rates as a function of time across neurons necessarily creates variability in the dictionaries across these neurons. Beyond this, we observe additional  variability among the dictionaries that is {\em not} explained by the rate fluctuations. For example, we can differentiate one of the four birds from two of the others just by looking at the proportions of high-order codewords (an average of $0.21$ bits in Jensen-Shannon divergence between the target bird and the rest), which means that we need around five independent samples/codewords to distinguish this bird from the others). Further, the mean inter-spike interval (ISI) for codewords is different from that of other words in the dictionaries, and this difference is also bird-dependent, see Fig.~\ref{f2}(E).

\begin{figure}
	   \includegraphics[width=0.9\textwidth]{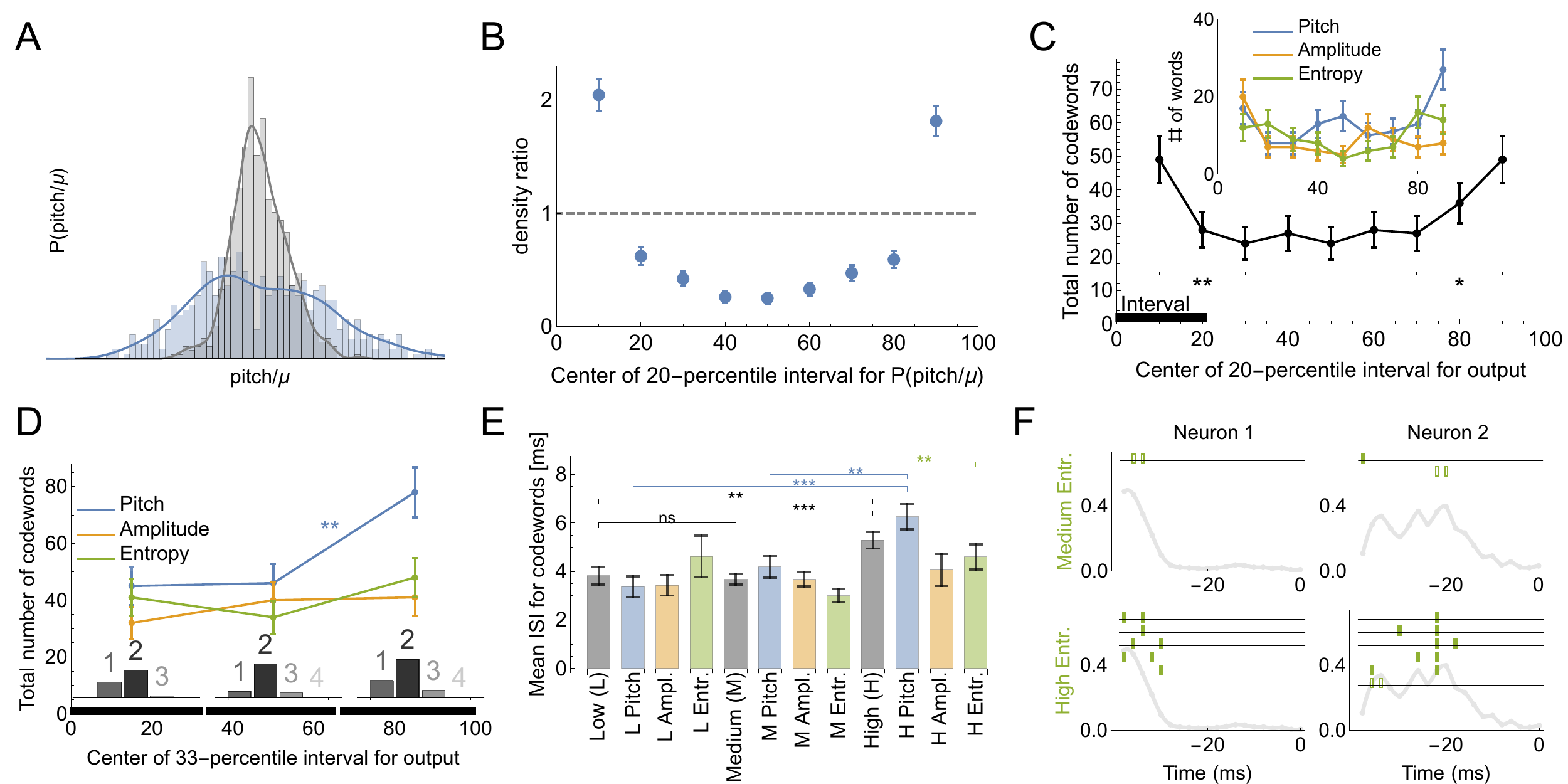}
\caption{\footnotesize{{\bf Codes for vocal motor exploration}. {\bf A:} Distribution of syllable pitch relative to the mean for exploratory and performance behaviors (blue, intact birds, vs.\ grey, LMAN-lesioned animals, see main text). {\bf B:}  Ratio of the histograms in (A) evaluated in the quintiles of the exploratory (blue) distribution centered around $[10\%, 20\%, \dots 90\%]$ points. {\bf C:} Total number of codewords when considering the vocal output as $1$ if it belongs to a specific $20$-percentile interval of the output distribution, and $0$ otherwise. We observe that there are significantly more codewords for the exploratory behavior (tails of the distribution compared to the middle intervals). Notice that the shape of the curves parallels that in (B), suggesting that exploration drives the diversity of the codewords. {\bf D:} Number of codewords when considering the vocal output as $1$ if it belongs to a $33$-percentile (non-overlapping) interval of the output distribution, and $0$ otherwise. Here there are significantly more codewords when coding for high pitch. Further, the codewords found for each of the three intervals are mostly multi-spike (histograms show the distribution of the order of the codewords for each percentile interval). {\bf E:} For codewords for the $33$-percentile intervals, we compare the mean inter-spike intervals (ISIs). Codewords for high outputs (especially for pitch and spectral entropy) have a significantly larger mean ISI. {\bf F}: We illustrate dictionaries of two neurons for the medium and the high spectral entropy ranges. Notice that the high entropy range has significantly more codewords.}}
\label{f3}
\end{figure}

\subsection*{Verification of the inferred dictionaries}

To show that the dictionaries we decoded are biologically (and not just statistically) significant, we verify whether the codewords can, in fact, be used to predict the behavioral features. For this, we built two logistic regression models that relate the neural activity to behavior. The first one uses the presence / absence of spikes in individual time bins and the second the presence / absence of the uBIA detected codewords as predictor variables (see {\em Online Methods}).  Note that the individual spikes model is still a precise-timing model, which has 20 predictors (20 time bins, each 2 msec long), and hence one may expect it to predict better than the codewords model, which typically has many fewer predictors. To account for the possibility of overfitting, in all comparisons we test the predictive power of models using cross-validation. We emphasize that we do not expect either of the two models to capture an especially large fraction of the behavioral variation. Indeed, Tang et al.~\cite{tang2014millisecond} have shown that, at 2 ms resolution, on average, there is only about 0.12 bits of information between the activity of an individual neuron and the behavioral features, and the assumption behind our entire approach is that none of individual predictors have strong effects. Further, a specific model, such as logistic regression, will likely recover even less predictive power from the data. With this, Supplementary Fig.~\ref{f6} compares prediction between the two models, obtaining a significantly higher accuracy and a lower mean cross-entropy between the model and the data for the models that use codewords as predictors. In other words, the higher order, but smaller, dictionaries decoded by uBIA outperform larger, non-specific dictionaries in predicting behavior.

\subsection*{Dictionaries for exploratory vs.\ typical behaviors}
Bengalese finches retain the ability to learn through their lifetimes, updating their vocalizations based on the sensorimotor feedback \cite{kuebrich2015variations,KellySober,SoberBrainard2009}. A key element of this lifelong learning capacity is the precise regulation of vocal variability, which songbirds use to explore the space of possible motor outputs, cf.~Fig.~\ref{f3}A,B. For example, male songbirds minimize variability when singing to females during courtship, but   dramatically increase the range of variability in acoustic features such as pitch when singing alone \cite{HesslerDoupe,woolley2014}. The variability is  controlled by the activity of nucleus LMAN. Silencing or lesioning LMAN reduces the acoustic variance of undirected song (Fig.~\ref{f3}A) to a level approximately equal to that of  female-directed song  \cite{kao2005contributions,olveczky2005}.  Using uBIA, we can ask for the first time whether the statistics of codewords controlling the exploratory vs.\ the baseline range of motor variability are different. To do this, we analyze the statistics of codewords representing different parts of the pitch distribution. First, we define the output as $\sigma_0=1$ if the behavior belongs to a specific $20$-percentile interval ($[0-20]$, $[10-30]$, \dots, $[80-100]$) and compare the dictionaries that code for behavior in each of the intervals. We find that there are significantly more codewords for exploratory behaviors (percentile intervals farthest from the median, cf.~Fig.~\ref{f3}C). This holds true for different features of the vocal output, though the results are only statistically significant if pooled over all features. To improve statistical power by increasing the number of trails in each acoustic interval, we also consider a division of the output into three equal intervals: low, medium, and high. In this case, there are still more codewords for the high exploratory pitch, and the dictionaries for each of the intervals are still multispike, cf.~Fig.~\ref{f3}D. We further observe that the codewords themselves are different for the three percentile groups: the mean ISI of high pitch, amplitude, and spectral entropy codewords is higher, with the largest effect coming from the pitch and the spectral entropy, cf.~Fig.~\ref{f3}E. Examples of typical and exploratory dictionaries are illustrated in Fig.~\ref{f3}F. 

These findings challenge common accounts of motor variability, in songbirds and other systems, that motor exploration is induced by adding random spiking variations to a baseline motor program. Rather, the over-abundance of codewords in the exploratory flanks of the acoustic distributions indicates that the mapping between the neural activity and the behavior is more reliable than in the bulk of the behavioral activity: multiple renditions of the same neural command result in the same behaviors more frequently, making it easier to detect the codewords. One possibility is that the motor system is less biomechanically noisy for large behavioral deviations. This  is unlikely due to the tremendous variation in the acoustic structure (pitch, etc.) of different song syllables within and across animals \cite{SoberBrainard2009,elemans2015}, which indicates that songbirds can produce a wide range of sounds and that particular pitches (i.e., those at at one syllable's exploratory tail) are not intrinsically different or harder for an animal to produce. Similarly, songbirds can dramatically modify syllable pitch in response to manipulations of auditory feedback \cite{SoberBrainard2009,kuebrich2015variations}. A more likely explanation for the greater prevalence of codewords in the exploratory tails is that the nervous system drives motor exploration by selectively introducing particular patterns into motor commands that are specifically chosen for their reliable neural-to-motor mapping. This would result in a more accurate deliberate exploration and evaluation of the sensory feedback signal, which, in turn, is likely to be useful during sensorimotor learning \cite{zhou2017chance}. 

\section*{Discussion}

In this work, we developed the unsupervised Bayesian Ising Approximation as a new method for reconstructing biological
dictionaries --- the sets of anomalously represented joint activities of multiple components of biological systems.  Inferring these dictionaries directly from data is a key problem in many fields of modern data-rich biological and complex systems research including systems biology, immunology, collective animal behavior, and population genetics.  
Our approach addresses crucial shortcomings that so far
have limited applicability of other methods. First, it does not limit  the possible  dictionaries, either by considering words of only limited length or of a pre-defined structure. Instead we performs a systematic analysis through all possible words that occur in the data sample. Second, it promotes construction of irreducible dictionaries, de-emphasizing related, co-occurring words. Further, uBIA does not make assumptions about the linear structure of dependencies unlike various linear methods.

To illustrate capabilities of the method, we applied it to analysis of
motor activity in cortical area RA in a songbird. We were able to
infer statistically significant codewords from large-dimensional
probability distributions ($2^{21}$ possible different words) with relatively
small data sets ($\sim 10^2\dots10^3$ samples). We verified that the codewords are biologically meaningful, in the sense that they predict behavioral features better than alternative approaches. Importantly, most of words in hundreds of dictionaries that
we reconstructed were more complex than is usually considered,
involving multiple spikes in precisely timed patterns. The
multi-spike, precisely timed nature of the codes was universal across
individuals, neurons, and acoustic features, while details of the codes (e.g., specific codewords and their number)
showed tremendous variability.

Further, we identified codewords that correlate with three different acoustic features of the behavior (pitch, amplitude, and spectral entropy), and different percentile ranges for each of these acoustic features. Across many of these analyses, various statistics of codewords predicting exploratory vs.\ typical behaviors were different. Specifically, the exploratory dictionaries contained more codewords than the dictionaries for typical behavior, suggesting that the exploratory spiking patterns are more consistently able to evoke particular behaviors. This is surprising since the exploratory behavior is usually viewed as being noisier than the typical one. Crucially, exploration is a fundamental aspect of sensorimotor learning  \cite{TumerBrainard2007,kuebrich2015variations,KellySober}, and it has been argued that
large deviations in behaviors are crucial to explaining the observed
learning phenomenology \cite{zhou2017chance}. However, the neural
basis for controlling exploration vs.\ typical performance is not well
understood. Intriguingly, vocal motor exploration in songbirds is driven by the output of a cortical-basal ganglia-thalamo-cortical circuit, and lesions of the output nucleus of this circuit (area LMAN) abolishes the exploratory (larger) pitch deviations \cite{kao2005contributions,olveczky2005}. Our findings therefore suggest that the careful selection of the spike patterns most consistently able to drive behavior may be a key function of basal ganglia circuits.

While the identified codewords are statistically significant, and we show that they can predict the behavior better than larger, but non-specific features of the neural activity, a
crucial future test of our findings will be in establishing their {\em
  causal} rather than merely correlative nature by means of
stimulating neurons with patterns of pulses mimicking the identified
codewords \cite{Srivastava31012017}. This will be facilitated by the
speed of our method, which can reconstruct dictionaries in real time
on a laptop computer. Additional future work can explore how
population-level dictionaries are built from the activity of individual
neurons, how the dictionaries develop and are modified in development, and whether the structure of dictionaries as a whole
can be predicted from various biomechanical and information-theoretic optimization principles. Finally, one needs to understand how these dictionaries are implemented in the recurrent dynamics of neural networks in
animals' brains. 

\section*{Online Methods}\label{methods}

\subsection*{Overview of neural decoding methods}\label{om:overview}

For many different experimental systems, it has been possible to measure the information content of spike trains \cite{fairhall2012information,tang2014millisecond,Srivastava31012017}, but the question of decoding -- which spike patterns carry this information? -- has turned out to be a harder one. Multiple approaches have been used to address this problem and to reconstruct neural dictionaries,
whether in the context of sensory or motor systems, starting with
linear decoding methods \cite{bialek1991reading}. All have fallen a
bit short, especially in the context of motor codes, where an animal
is free to perform any one of many behaviors it wishes, and hence
statistics are usually poor. A leading method is Generalized Linear
Models (GLMs)
\cite{paninski2004maximum, pillow2008spatio, gerwinn2010bayesian},
which encode the rate of spike generation from a certain neuron at a
certain time as a nonlinear function of a linear combination of past
stimuli (sensory systems) or of future motor behavior (motor systems)
and the past spiking activity of a neuron and its presynaptic
partners. GLM approaches can detect the importance of the timing of
individual spikes and sometimes interspike intervals for information
encoding, but generalizations to detect importance of higher order
spiking patters are not yet well established. Another common approach
is based on maximum entropy (MaxEnt) models
\cite{schneidman2006weak,GranotAtedgi:2013cwb,Savin:2017jc}. These
replace the true distribution of the data with the least constrained
(i.~e., maximum entropy) approximation consistent with low-order,
well-sampled correlation functions of the distribution. The approach
is computationally intensive, especially when higher order
correlations are constrained by data. At the same time, to approximate
empirical distributions well, a large number of such constraints is
often required. This needs very large datasets, especially if one is
interested in relating the neural activity to the external (behavioral
or sensory) signals. Such large datasets are hard to obtain in the
motor control setting. More recently, feed-forward and recurrent
artificial neural network approaches have been used to decode
large-scale neural activity \cite{pandarinath2018inferring, Glaser:2017we},
but these have focused primarily on neural firing rates over large
(tens of milliseconds) temporal windows. As a result, to date, there
have not been successful attempts to reconstruct neural dictionaries
from data, which would (i) resolve spike timing in words of the
dictionary to a high temporal resolution, (ii) be comprehensive and
orthographically complex, not limiting the words to just single spikes
or pairs of spikes, and (iii) discount correlations among spiking
words to produce irreducible dictionaries that only detect those
codewords that cannot be explained away by correlations with other
words in the dictionary. 

\subsection*{Details of the Bayesian Ising Approximation approach}\label{om:details}

\begin{figure}
\includegraphics[width=0.9\textwidth]{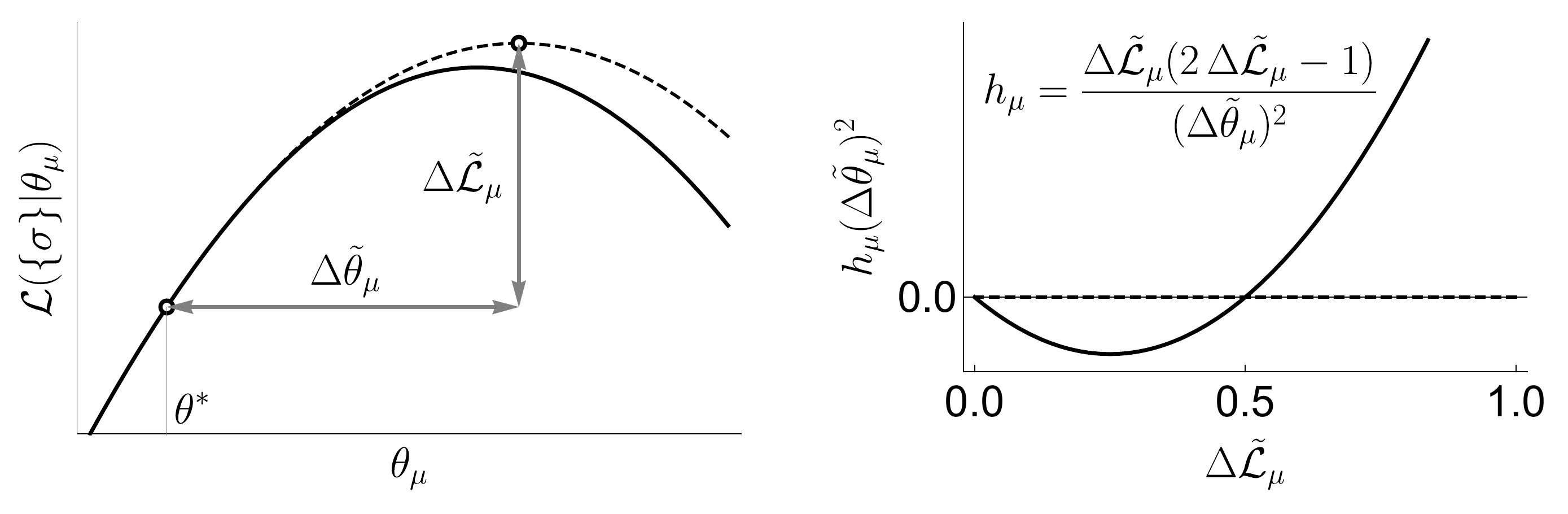}
\caption{\footnotesize{ {\bf Geometric interpretation of the fields $h_{\mu}$ in the uBIA method, in relation to the log-likelihood function $\mathcal{L}(\{\bm{\sigma}\}|\bf{\theta})$.} The uBIA method makes an approximate guess (dashed line in left panel) of how much in log-likelihood $\Delta \mathcal{\tilde{L}}_{\mu}$ we would win be fitting a parameter $\theta_{\mu}$, and how far in parameter space we would need to go, $\Delta \tilde{\theta}$ (see left panel). The sign of the field only depends on the improvement in log-likelihood, being positive beyond a threshold (inclusion of a word). This complexity penalization comes from the Bayesian approach in this strong regularized regime. On the other hand, the farther we go in parameter space, the smaller in absolute value the field becomes (see right panel).}}
\label{fa1}
\end{figure}

To obtain Eq.~(\ref{s2e1}) for the posterior probability of including
a word into the dictionary, we start with
\begin{equation}
P(\mathbf{s}|\bs)\propto p(\bs|\mathbf{s}) = 
			\int \mathbf{d}\bt\, p(\{\bs|\bt,\mathbf{s}) \prod_{\mu \mid s_{\mu}=1} p_{\epsilon}(\theta_{\mu}|s_{\mu}).
\label{eq01}
\end{equation}
Now we make two approximations. First, we
evaluate the integral in Eq.~(\ref{eq01}) using the saddle point
approximation around the peak of the {\em prior}, $\bm{\theta}^*$. This
is a low signal-to-noise limit, and it is different from most
high signal-to-noise approaches that analyze the saddle around the peak of the {\em posterior}.  Second, we do all
calculations as a Taylor series in the small parameter $\epsilon$ (see
below on the choice of $\epsilon$). Both approximations are facets of
the same strong regularization assumption, which insists that most
coupling constants $\theta_\mu$ are small. Following Fisher and Mehta \cite{fisher2015bayesian}, we obtain Eq.~(\ref{s2e1}), where the
magnetic fields (biases) $h_{\mu}$ and the exchange interactions
$J_{\mu\nu}$ are
\begin{equation}
\begin{array}{ll}
h_{\mu}(\bs) =&\displaystyle \frac{1}{2} \left. \left[ \frac{\partial^2 \mathcal{L}}{\partial \theta_{\mu}^2}+ \left( \frac{\partial \mathcal{L}}{\partial \theta_{\mu}}\right)^2\right] \right|_{\bm{\theta}^*}, \\\\
J_{\mu\nu}(\bs) =&\displaystyle \frac{1}{4} \left. \left[\frac{\partial^2 \mathcal{L}}{\partial \theta_{\mu}\partial \theta_{\nu}} \left(\frac{\partial^2 \mathcal{L}}{\partial \theta_{\mu}\partial \theta_{\nu}}+ 2 \frac{\partial \mathcal{L}}{\partial \theta_{\mu}}\frac{\partial \mathcal{L}}{\partial \theta_{\nu}}\right)\right]\right|_{\bm{\theta}^*},
\end{array}
\label{eq02}
\end{equation}
where $\mathcal{L}=\log P(\bs| \bt)$ is the log-likelihood (see Fig.~\ref{fa1} for a geometric interpretation of the field $h_{\mu}$). Plugging
in the  model of the probability distribution, Eq.~(\ref{loglinear}),
we get for the fields and the exchange interactions 
\begin{align}
h_{\mu} &= \frac{M^2}{2}\left[ (\overline{\bs}_{\mu}- \langle \bs_{\mu} \rangle)^2-\frac{\mathrm{var}(\bs_{\mu})}{M} \right],\label{field}
\\
J_{\mu\nu} &=\frac{M^2}{4}  \mathrm{cov}(\bs_{\mu},\bs_{\nu}) \left[\mathrm{cov}(\bs_{\mu},\bs_{\nu})-2 M(\overline{\bs}_{\mu}- \langle \bs_{\mu} \rangle) (\overline{\bs}_{\nu}- \langle \bs_{\nu} \rangle) \right].\label{exchange}
\end{align}
Here, to simplify the notation, we defined
$\bs_{\mu}\equiv \prod_{i \in V_{\mu}} \sigma_i$. Further, angular
brackets, $\mathrm{cov}$, and $\mathrm{var}$ denote the {\em a priori}
expectations, covariances, and variances of frequencies of words in
the null model, which matches frequency of occurrence of each
individual $\sigma_i$ (probability of firing in every time bin for the
songbird data). Similarly, overlines denote the empirical counts or
correlations between co-occurrences of words in the observed
data. Specifically, denoting by $n_{\mu}$ the marginal
frequencies of the word $V_{\mu}$ in the data, these expectations and
frequencies are defined as follows:
\begin{align}
 \overline{\bs}_{\mu} &= \frac{1}{M}\sum_{m=1}^M \left(\prod_{i \in V_{\mu}} \sigma_i^{\{m\}} \right)= \frac{n_{\mu}}{M},\\
\langle \bs_{\mu} \rangle &= \prod_{i \in V_{\mu}} \langle \sigma_i \rangle = \prod_{i \in V_{\mu}} \frac{n_i}{M}, \label{mean}\\
\mathrm{var}(\bs_{\mu}) &= \langle \bs_{\mu}^2 \rangle - \langle \bs_{\mu} \rangle^2 = \langle \bs_{\mu} \rangle \left(1 - \langle \bs_{\mu} \rangle \right)=\prod_{i \in V_{\mu}} \frac{n_i}{M} \left(1-\prod_{i \in V_{\mu}} \frac{n_i}{M}\right),\\
\displaystyle \mathrm{cov}(\bs_{\mu},\bs_{\nu}) &= \langle\bs_{\mu}\bs_{\nu}\rangle -\langle\bs_{\mu}\rangle\langle\bs_{\nu}\rangle = \prod_{k \in V_{\mu}\cup V_{\nu}} \frac{n_k}{M} -\left(\prod_{i \in V_{\mu}} \frac{n_i}{M}\right)\left(\prod_{j \in V_{\nu}} \frac{n_j}{M}\right),
\end{align}
To derive these equations, note that $\sigma_i^2=\sigma_i$. Note also
that $\mathrm{cov}(\bs_{\mu},\bs_{\nu})=0$ if the intersection of
$V_{\mu}$ and $V_{\nu}$ is empty. 

Equation~(\ref{field}) has a straightforward interpretation, which we illustrate in Fig.~\ref{fa1}. Specifically, if the
difference between the {\em a priori} expected frequency and the
empirical frequency of a word is statistically significantly nonzero
(compared to the {\em a priori} standard error), then the
corresponding word is anomalously represented.  It does not matter
whether the word is over- or under-represented: in either case, if the
frequency deviated from the expectation, then the field $h_{\mu}$ is
positive, biasing the indicator $s_\mu$ towards 1, and hence towards
inclusion of the word in the dictionary. If the frequency is as
expected, then the field is negative, and the indicator is biased
towards 0, excluding the word from the dictionary. Note that as $M$ increases, the standard error goes down, and the field generally increases, allowing us to consider more words. The sign of
$\theta_{\mu}$ would determine whether the word is over- or
underrepresented. However, estimating the exact value of $\theta_\mu$
from small datasets is often impossible and is not our goal, even
though, in Fig.~\ref{f1}, we denote words as under- or
over-represented by whether their empirical frequency is smaller or
larger than the {\em a priori} expectation. Thus in some aspects, our
approach is similar to the previous work \cite{schnitzer2003multineuronal}, where
multi-neuronal patterns are found by comparing empirical firing
probabilities to expectations. However, we do this comprehensively for
{\em all} patterns that occur in data, and we account for reducibility
of the dictionaries (also see below).

The exchange interactions $J_{\mu\nu}$ are also interpretable. As
explained above, correlations among words are a serious
problem. Indeed, for example, a word $\sigma_0\sigma_i\sigma_j = 1$
may occur too frequently simply because its sub-words
$\sigma_0\sigma_i = 1$, $\sigma_i\sigma_j = 1$, or
$\sigma_0\sigma_j = 1$ are common. Alternatively, the word may be
frequent because, in its turn, it is a sub-word of a larger common
word, for example, $\sigma_0\sigma_i\sigma_j\sigma_k = 1$. In GLMs,
resolving these overlaps requires imposing sparsity or other
additional constraints. In contrast, the couplings $J_{\mu\nu}$
address this problem for uBIA naturally and computationally
efficiently. Notice that because of the factor 2 in
Eq.~(\ref{exchange}), the exchange interactions are predominantly
negative if one expects the two studied words to be correlated, and if
they co-occur in the empirical data as much as they are expected to
co-occur in the null model because of the overlaps in their
composition, $V_{\mu}$ and $V_{\nu}$. Negative $J_{\mu\nu}$s implement
a ``winner-take-all'' mechanism, where statistical anomalies in data
that can be explained, in principle, by many different $\theta_{\mu}$s
are attributed predominantly to one such $\theta_\mu$ that explains them
the best. On the other hand, the exchange interactions are positive if
one expects correlations between the words {\em a priori}, but does not
observe them. Thus, in principle, a word can be included in the
dictionary even at zero field $h_\mu$.

Knowing the coefficients $h_{\mu}$ and $J_{\mu\nu}$, one can
numerically estimate $\langle s_{\mu} \rangle$, the posterior
expectation for including a word $V_{\mu}$ in the
dictionary. Generally, finding such marginal expectations from the
joint distribution in disordered systems is a hard problem. However,
here $h_{\mu} \propto \epsilon$ and $J_{\mu\nu} \propto \epsilon^2$,
so that the fields and the interactions create small perturbations
around the ``total ignorance'' solution,
$\langle s_{\mu} \rangle=1/2$ (this is a manifestation of our general assumption that none of the words is very easy to detect). Therefore, we calculate the marginal
expectation using fast mean field techniques \cite{opper2001advanced}. We use the {\em
  naive} mean field approximation, which is given by  self-consistent
equations for the posterior expectations in terms of the magnetizations
$m_{\mu}=2 \langle s_{\mu} \rangle-1$,
\begin{align}
\tanh^{-1}(m_{\mu}(\epsilon)) &= \frac{\epsilon}{2} \left[ h_{\mu} +\epsilon \sum_{\nu} J_{\mu\nu} + \frac{\epsilon}{2} \sum_{\nu} J_{\mu\nu} m_{\nu}(\epsilon) \right] \label{meanfield}\\
&= \frac{1}{2} \left[\epsilon h_{\mu} + \epsilon^2 h^{\text{eff}}_{\mu}(\epsilon) \right],\label{fieldeff}
\end{align}
so that interactions among spins are encapsulated in an effective
field $\epsilon h^{\text{eff}}_{\mu}$. We solve Eq.~(\ref{meanfield})
iteratively \cite{fisher2015bayesian}, by increasing $\epsilon$ from
0 ---that is, from the total ignorance $\langle s_{\mu} \rangle=1/2$
or $m_{\mu}=0$ --- and up to the limiting value
$\epsilon_{\text{max}}$ in steps of $\delta \epsilon=M^{-1}/20$. This
limiting value $\epsilon_{\rm max}$ is determined by the two
approximations involved in the strong regularization
assumption. First, the saddle point approximation around the peak of
the prior in Eq.~(\ref{eq01}) implies that the characteristic width of
the prior should be smaller than that of the likelihood,
$\epsilon \le \epsilon_1=1/M$. Second, the Taylor series up to second
order in $\epsilon$ for the posterior of the indicator variables
implies that the quadratic corrections should not be larger than the
linear terms. Within the mean field approximation, this means that
$\langle|h_{\mu}|\rangle_\mu \ge \langle|\epsilon
h_{\mu}^{\text{eff}}(\epsilon)|\rangle_\mu$, which is saturated at
some $\epsilon_2$ (notice that, in contrast to our usual notation, the
averages here are over the indices, and not the data).  Thus overall
we take $\epsilon_{\rm max}=\text{min}\{\epsilon_1,\epsilon_2\}$.

Additionally we have used the TAP equations \cite{opper2001advanced}, instead of Eq.~(\ref{meanfield}) to calculate magnetizations. These are more accurate since they account for how a spin affects itself through its couplings with the other spins. However, corrections due to this more complicated method were observed to be negligible in our strong regularized regime, since they were of higher order in $\epsilon \ll 1$. Thus all results that we report are based on the mean field estimation.

\subsubsection*{Effect of absent words}\label{om:absent}
Of the exponentially many possible words, majority will not happen in a realistic data set. In particular, this includes most of long words. At the same time, {\em a priori} expectations for the frequency of such words, Eq.~(\ref{mean}), decrease exponentially fast with the word length. Thus the fields, Eq.~(\ref{field}), for the words that do not occur are small, and the posterior expectation for including these words in the dictionary is $\langle s_\mu\rangle\approx 1/2$, so that we do not need to analyze them explicitly. A bit more complicated is the fact that all words affect each other's probability to be included in the dictionary through the exchange couplings $J_{\mu\nu}$, so that, in principle, the sum in the mean field equations, Eq.~(\ref{fieldeff}), is over exponentially many terms. Here we show that the effect of non-occurring words on the interaction terms is exponentially small in $N$, as long as the empirical averages $n_i/M\ll 1$. 

To illustrate this, we start with the probabilities $p(\sigma_i=1)=p_i$  of a single variable $i$ being active. We then define the average such probability $q=N^{-1}\sum_i p_i$. Without the loss of generality, we assume $q<1/2$, and otherwise we rename $\sigma_i\to1-\sigma_i$.  Denoting a long word of a high order $k$ that does not occur in the data as $\bs_\omega$, we have $n_{\omega}=0$. Then the corresponding field is 
\begin{align}
h_{\omega} &= \frac{M^2}{2}\left[ (0 - \langle \bs_{\omega} \rangle)^2-\frac{\mathrm{var}(\bs_{\omega})}{M} \right] \label{fieldhigh0}\\
&\sim - \frac{M}{4} q^k \left[1-q^k M \right] \sim - \frac{M}{4} q^k.\label{fieldhigh}
\end{align}
Here we consider as {\it high order} words those, for which $q^k M\ll 1$ (in general, $\langle\sigma_{\omega} \rangle M=\langle n_{\omega} \rangle \ll 1$, which happens for $k\sim 4\dots5$ for our datasets). Then the  magnetization is 
\begin{align}
m_{\omega}(k) &\simeq \tanh\left(\frac{\epsilon}{2}h_{\omega} \right) \sim -\frac{\epsilon M}{8} q^k . \label{magnhigh}
\end{align}
This illustrates our first assertion that none of these non-occurring words will be included in the dictionary.  However, as a group, they may still have an effect on words of lower orders. To estimate this effect, for a word $\bs_\mu$ of a low order $k_0$, we calculate the effective field $\tilde{h}^{\text{eff}}_\mu$, which all of the non-occurring words $\bs_\omega$ have on it. First we notice that, if $V_{\mu}$ and $V_{\omega}$ do not overlap, then their covariance is zero, and $J_{\mu \omega}=0$. That is, only high order words that overlap with $V_{\mu}$ can contribute to $\tilde{h}^{\text{eff}}_\mu$. 
Since  $\mathrm{cov}(\bs_{\mu},\bs_{\omega}) \sim q^k(1-q^{k_0})$, the couplings are
\begin{equation}
J_{\mu\omega}(k)  \sim \frac{M^2}{4} q^{2k}(1-q^{k_0}) \times \mathcal{O}(1).
\end{equation}
Using Eq.~(\ref{meanfield}), this gives for the typical effective field that absent words have on the word $\mu$
\begin{align}
\tilde{h}^{\text{eff}}_\mu(\{\omega\}\rightarrow \mu) &\simeq \epsilon \sum_{k \gtrapprox k_0}^N \mathcal{N}(k)\, J_{\mu\omega}(k)\,(1+\frac{\epsilon}{2} m_{\omega}(k)),\label{heffhigh2}
\end{align}
where the number of words of order $k$ that overlap with $\bs_\mu$ and can affect it is given by the combinatorial coefficient
$\mathcal{N}(k) \simeq \binom{N-k_0}{k-k_0}$. This  has a very sharp peak at $k=(N+k_0)/2$, where $\mathcal{N}\simeq 2^{N-k_0}$. We can approximate the sum in Eq.~(\ref{heffhigh2}) as the argument of the sum evaluated at this peak $k=(N+k_0)/2$, obtaining an effective field coming from high order words
\begin{align}
\tilde{h}^{\text{eff}}_\mu (\{\omega\}\rightarrow \mu) &\sim \epsilon\, 2^{N-k_0} \frac{M^2}{4} q^{N+k_0}(1-q^{k_0}) \left[ 1-\frac{\epsilon^2 M}{16} q^{\frac{N+k_0}{2}}\right]\\
&\sim \frac{\epsilon\,M^2}{4} \left(\frac{q}{1/2}\right)^N {\frac{q}{2}}^{k_0}(1-q^{k_0})\\
&\propto \left(\frac{q}{1/2}\right)^N.
\end{align}
In other words, even the combined effect of all higher order absent words is small if the average frequency of individual letters is smaller than 1/2. We thus can disregard all non-occurring words in the mean field equations.

We stress that, for this to hold, the average of the binary variables $\sigma_i$ must be small,  $q=N^{-1}\sum_i p(\sigma_i=1)<1/2$. In our songbird dataset, this condition was fulfilled with $q \sim 0.2$. However, in $4\%$ of cases the probability to have a spike in a certain time bin was $p_i>1/2$. Thus to stay on the safe side, we performed additional analyses by redefining variables as $\sigma_i\to 1-\sigma_i$ if the presence of a spike in a bin was $>50\%$. In other words, in such cases, we defined the absence of the spike as 1 and the presence as 0. For our datasets, the findings did not change with this redefinition.

This previous analysis does not imply that absent words of high order are irrelevant --- it only says that they cannot be detected with the available datasets. In the numerical implementation of the method, we filter out long absent words $\omega$ such that $\langle \sigma_{\omega} \rangle M =\langle n_{\omega} \rangle  <0.02$, with this cutoff determined by Eqs.~(\ref{fieldhigh0}-\ref{magnhigh}), so that, for these words, $h_{\omega}\ll 1$. These words get assigned 1/2 as the posterior probability of inclusion in the dictionary, and their contribution to the mean field equations is neglected. In contrast, if a word $\omega$ is absent but $\langle n_{\omega} \rangle  \ge 0.02$, we include them in the analysis, Eq.~(\ref{s2e1}). Such words may  turn out to be relevant code words, especially if they happen a lot less frequently than expected {\em a priori}.

\subsection*{Synthetic data for testing and fine-tuning the method}\label{om:synthetic}

To set the free parameters of our approach and to quantify its ability to reconstruct dictionaries, we test it on synthetic data sets that are similar to the songbird data, analyzed in this work. We use the log-linear model, Eq.~(\ref{loglinear}), as a generative model for binary correlated observables $\bs$ with $N=20$. We choose the individual biases in the generative model from a normal distribution, $\theta_i \sim \mathcal{N}(-0.7,0.1^2)$, which matches the observed probability of a spike in a bin in the bird data. That is, $p(\sigma_i=1) \simeq [1+\exp(-2\theta_i)]^{-1} \sim q \sim 0.2$. Then we select which binary variables interact. We allow interactions of 2nd, 3rd, and 4th order, with an equal number of interactions per order. For different tests, we choose the interaction strengths from (a) the sum of two Gaussian distributions, one with a positive mean and the other with a negative one, ${\rm mean}({\theta}_\mu)= \pm 0.5$, ${\rm std}(\theta_\mu)=0.1$, and (b) from one Gaussian distribution centered at zero with ${\rm std} (\theta_\mu)=0.5$. Both  choices reflect our strong regularization assumption, so that effects of  individual variables on each other are weak, and a state of one variable does not determine the state of the others, and hence does not ``freeze'' the system. We are specifically interested in performance of the algorithm in the case where $\bt$ are distributed as the sum of Gaussians. On the one hand, this tests how the algorithm deals with data that are atypical within its own assumptions. On the other hand, this choice ensures that there are very few values of $\bt$ that are statistically indistinguishable from zero, making it easier to quantify findings of the algorithm as either true or false. We have additionally tested other distributions of $\bt$, but no new qualitatively different behaviors were observed. Finally, for both types of distributions of $\bt$, we also varied the density of interactions $\alpha$, from $\alpha=2$ to $\alpha=4$, which spans the interaction densities of tree-like and 2D lattice-like networks. 

Next, we generate $M$ samples from these random probability distribution and we apply our pipeline to reconstruct the dictionary. We test on $400$ distributions from each family. As the first step, we discard high-order words absent in the data using a threshold on the expected number of occurrences $\langle \sigma_{\mu}\rangle M=\langle n_{\mu}\rangle <0.02$, as explained above. Next, we select $N_{\text{max}}$ words that have the highest (absolute) values in magnetic field (we have tested $N_{\text{max}}=200,\,500,\,2000,\,5000$, and finally use $500$ after not observing differences). To decide which of these remaining words are to be included in the dictionary, we build the Ising model on the indicator variables, Eq.~(\ref{s2e1}), with its corresponding magnetizations $m_{\mu}$ given by the mean field equations. We start from an inverse regularization strength of $\epsilon=0$ and then increase $\epsilon$ in steps of $\delta\epsilon= 1/(20M)$, up to $\epsilon_{\rm max}=\text{min}\{\epsilon_1,\epsilon_2\}$, as detailed above. We then set the full false discovery rate $n_{\rm false}$ --- the number of dictionary words that we allow ourselves to identify falsely from the fully reshuffled data, which must have zero words in it.  Note that we do reshuffling while keeping the observed frequency of individual variables $n_i$ constant. For the value of $n_{\rm false}$, we identify the significance threshold for the magnetization, which allow this many false words to be accepted in the shuffled data, on average (from the same $400$ distributions). Finally, we select as dictionary words those that have their marginal magnetizations $m_{\mu}$ above a significance threshold $m_{\mu}>m(n_{\rm false})$. 

\begin{figure}
    \includegraphics[width=0.9\textwidth]{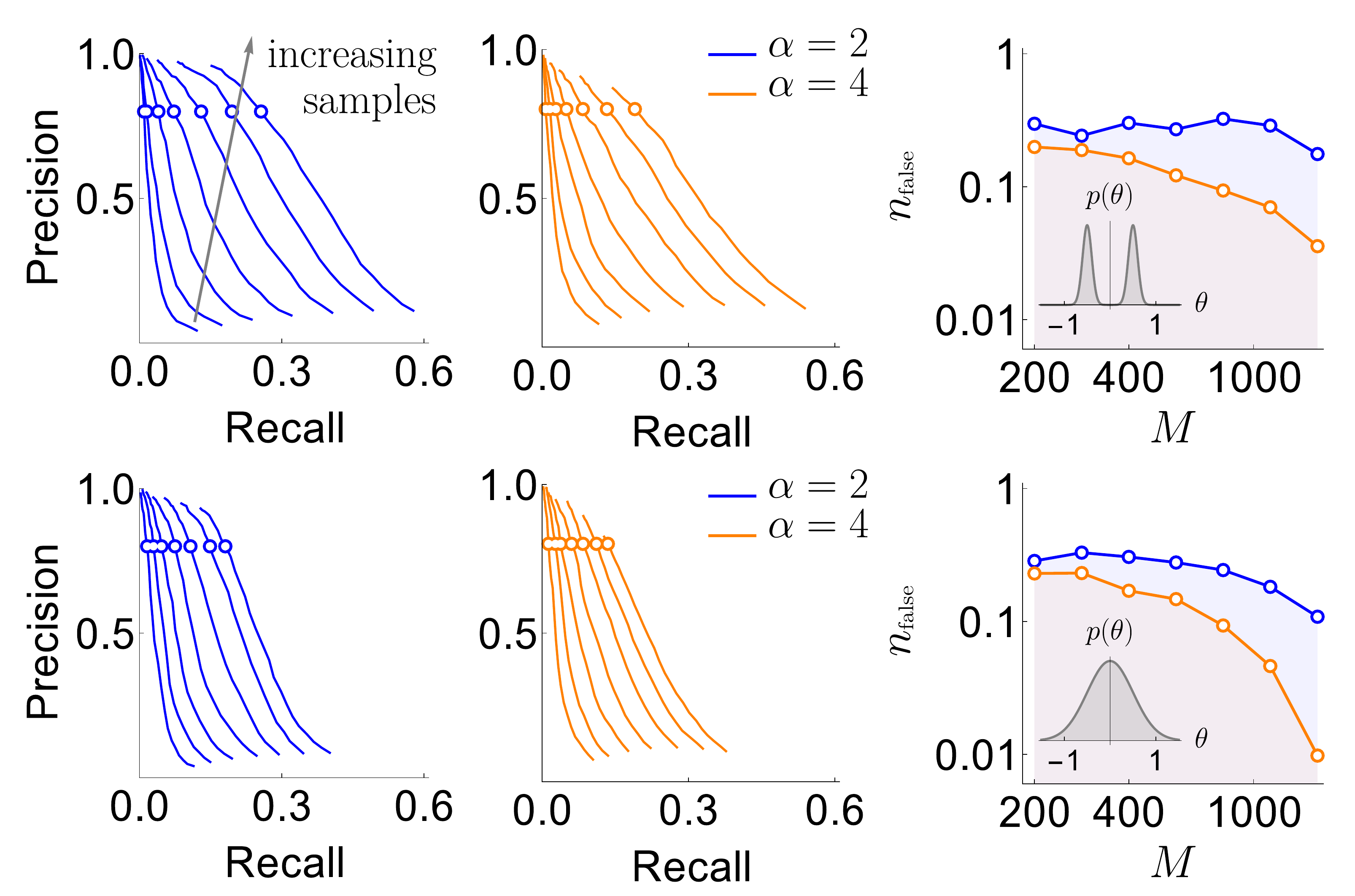}
\caption{\footnotesize{{\bf Results of the synthetic data analysis.} Performance on synthetic data as a function of the density of interactions $\alpha$, the distributions $p(\theta)$ for the strength of interactions, and the number of samples $M \in \{200,\dots,1600\}$ (logarithmically spaced). The first and the second columns correspond to precision-recall curves for the different density of interactions (significant words) per variable, $\alpha \in \{2,4\}$, within the true generative model. The top and the bottom rows corresponds to the interaction strengths $\bt$ selected from the sum of two Gaussian distributions, or a single Gaussian, as described in the text. For the first two columns, we vary the significance threshold in marginal magnetization $m(n_{\rm false})$, such that the full false discovery rate on the shuffled data  $n_{\rm false} \in [0.005,40]$.  In the third column we show the value of $n_{\rm false}$ that corresponds to the precision of $80\%$ as a function of $M$, so that the precision is larger than $80\%$ in the shaded region. This region is quite large and overlaps considerably for the four cases analyzed, illustrating robustness of the method. }}
\label{f5}
\end{figure}

We then repeat the analysis for different values of $n_{\rm false}$, seeking the thresholds that minimize false negatives (exclusions of true words from the dictionary) and false positives (inclusion of spurious words into the dictionary) simultaneously. To measure these, we use two metrics: precision and recall. First, precision measures the
fraction of the words included in the dictionary that are true, i.~e., have a nonzero $\theta_\mu$ in the generative model.  Second, recall measures the fraction of the words in the generative model with $\theta_\mu\neq 0$ that were included in the dictionary. Results of the analysis are shown in Fig.~\ref{f5}B. Since data set sizes are relatively small, we do not expect to detect all words, especially in the case where $\theta_\mu$ are allowed to be close to 0 in the generative model (Gaussian distributed). Thus we emphasize precision over recall in setting parameters of the algorithm: we are willing to not include words in a dictionary, but those words that we include should have a large probability of being true words in the underlying model.  Our most crucial observation is that the precision-recall curves are remarkably stable with the changing density of interactions. Recall is smaller when
interactions coefficients are taken from a Gaussian centered at zero. However, one could argue that missing words with very small $\theta_\mu$ should not be considered a mistake: they are not significant words in the studied dictionary. We observe that by keeping the full false discovery rate $n_{\rm false}$
below $0.5$ (only about half a word detected falsely, on average, in shuffled data), we can reach a precision as high as $80\%$, extracting
$20\%-30\%$ of the codewords depending on the number of samples, the distribution of $\bt$, and $\alpha$. We are thus confident that our method produces dictionaries, in which a
substantial fraction of words correspond to true words in the data.

\subsection*{Testing the predictive power of the uBIA dictionaries}

In this section, we test whether the codewords found in data from songbird premotor neurons can be used to predict the subsequent behavior. We compare two logistic regression models: one that uses the activity in the 20 time bins to predict the behavior and another that only uses as features the activity of the few relevant codewords, usually far fewer than 20. The features corresponding to the codewords are binary, and they are only active when all the time bins of such words are active. This means that the model using the time bins is more complex, as it already has all the information that the codewords model has and more, though it does not account for combinatorial effects of combining spikes into patterns. In order to properly test the predictive power between these two models with different complexity we perform 2-fold cross-validation, using a log-likelihood loss function. As is common in these cases, an L2 penalty is included to help the convergence to a solution (the models were implemented with the Classify function from Mathematica, whose optimization is done by the LBFGS algorithm). As shown by Tang et al.\cite{tang2014millisecond}, not all neurons in our dataset are timing neurons, or even code for the behavior at all. Thus we restrict the comparison to those cases that have at least 4 codewords (27 case  in total, with 10 codewords on average). Both of the logistic regression models have the following structure
\begin{equation}
\displaystyle p(y=1|\bm{z},\bm{\beta})=\frac{1}{1+\exp(-\beta_0-\sum_i \beta_i z_i)},
\label{om:logreg}
\end{equation}
where $y$ corresponds to the behavior, and the features correspond to the time bins in one case ($z_i=x_i$) and to the codewords in the other ($z_i=\prod_{j\in V_i} x_j$), while $\beta_i$ are the coefficients of the model. The loss function used is the log-likelihood with the L2 penalty,
\begin{equation}
\displaystyle \mathcal{L}= \sum_{m=1}^M \log p(y^{(m)}|\bm{z}^{(m)},\bm{\beta}) -\frac{\lambda}{2} \sum_i \beta_i^2,
\label{om:logregloss}
\end{equation}
where $M$ is the number of samples, and $\lambda$ is the regularization strength. In our analysis, as different datasets have different number of samples, we show the results for the mean cross-entropy over the test data, which correspond to the normalized log-likelihood.

\begin{figure}
    \includegraphics[width=0.8\textwidth]{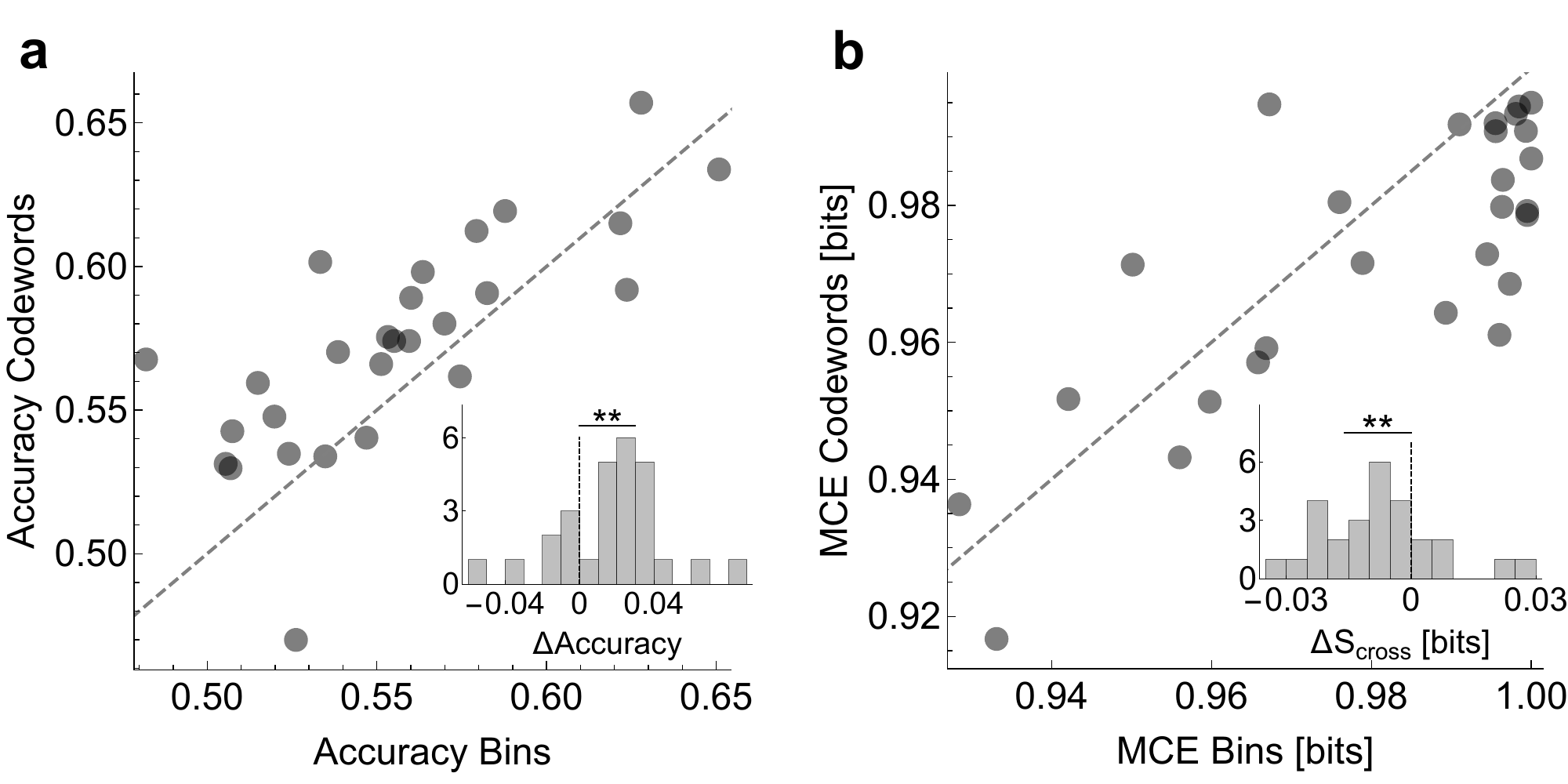}
\caption{\footnotesize{{\bf Prediction accuracy with uBIA dictionaries.} We compare prediction of the behavior using logistic regression models that have as features (i) neural activity in all the time bins at 2 ms resolution  versus (ii) only the detected relevant codewords.  {\bf A}: Scatter plot of accuracy of models of both types, evaluated using 2-fold cross-validation. Inset shows that the different between the prediction is significant with $p<0.01$ according to the paired t-test. {\bf B}: Scatter plots of the mean cross-entropy between the data and the models for the two model classes. Inset: Even though the models that use the codewords are simpler (have fewer terms), they are able to predict better (with lower cross-entropy) according to the paired t-test.}}
\label{f6}
\end{figure}

Tang el al.~\cite{tang2014millisecond} showed that individual neurons on average carry around $0.12$ bits at a 2 ms scale. So for both models we expect the prediction accuracy to be barely above chance, especially since we are focusing on a particular prediction model (a logistic regression), and may be missing predictive features not easily incorporated in it. Figure~\ref{f6}{\bf a} shows the scatter plot of accuracy in the 27 analyzed datasets, plotting the prediction using the time bins activity on the horizontal axis versus prediction using only the codewords activity on the vertical one. We observe that the models based on codewords are consistently better than the ones using all the 20 time bins, and the difference is significant (inset). We additionally evaluate the quality of prediction using the mean cross-entropy between the model and the data. Figure~\ref{f6}{\bf b} shows that the models with the codewords have lower mean cross-entropies and thus generalize better (see Inset). 

\section*{Software Implementation}
The software implementation of uBIA is available from \url{https://github.com/dghernandez/decomotor}.

\section*{Data}
The data used in this work is available from \url{https://figshare.com/articles/Songbird_premotor_dictionaries/10315844}.

\bibliography{sample}

\section*{Acknowledgements}

We thank David Hoffman and Pankaj Mehta for valuable discussions. This work was supported in part by NIH Grants R01-EB022872, R01-NS084844, and
R01-NS099375, and NSF grant BCS-1822677. IN acknowledges hospitality of the Kavli Institute for
Theoretical Physics, supported in part by NSF Grant PHY-1748958, NIH
Grant R25GM067110, and the Gordon and Betty Moore Foundation Grant
2919.01. IN and SJS further acknowledge hospitality of the Aspen
Center for Physics, which is supported by NSF
grant PHY-1607611.

\section*{Author contributions statement}

DGH, SJS, and IN conceived the project, performed the research, and wrote the paper. DGH wrote the software implementation of uBIA.

\section*{Additional information}
\textbf{Competing interests} The authors declare no competing interests.

\end{document}